\documentstyle{article}
\def\rd{{\rm d}}
\def\rD{{\rm D}}
\def\be{\begin{equation}}
\def\ee{\end{equation}}
\def\bea{\begin{eqnarray}}
\def\eea{\end{eqnarray}}

\begin{document}

\hfill{}

\vskip 3\baselineskip \noindent{\Large\bf Large-$N$ limit of the
non-local 2D Yang-Mills and generalized Yang-Milss theories on a cylinder}

\vskip \baselineskip

Khaled Saaidi$^{\; a,1}$ and Mohammad Khorrami$^{\; b,2}$

\vskip\baselineskip

\begin{itemize}
\item[{\small a:}]{\small Department of Physics, Tehran University,
North-Kargar Ave., Tehran, Iran}
\item[{\small b:}] {\small Institute for Advanced Studies in Basic Sciences,
P. O. Box 159, Zanjan 45195, Iran}
\item[{\small 1:}] {\small lkhled@molavi.ut.ac.ir}
\item[{\small 2:}] {\small mamwad@iasbs.ac.ir}
\end{itemize}

\vskip 2\baselineskip



\vskip 2\baselineskip

\begin{abstract}
The large-group behavior of the nonlocal YM$_2$'s and gYM$_2$'s on a
cylinder or a disk is investigated. It is shown that this behavior is
similar to that of the corresponding local theory, but with the area of the
cylinder replaced by an effective area depending on the dominant
representation. The critical areas for nonlocal YM$_2$'s on a cylinder with
some special bounary conditions are also obtained.
\end{abstract}

\newpage
\section{Introduction}
Pure two-dimensional Yang-Mills theories (YM$_2$) have certain
properties, such as invariance under area-preserving
diffeomorphisms and lack of any propagating degrees of freedom.
There are, however, ways to generalize these theories without
losing these properties. One way is the so-called generalized
Yang-Mills theories (gYM$_2$'s). In a YM$_2$, one starts from a
$B$-$F$ theory in which a Lagrangian of the form $i{\rm
tr}(BF)+{\rm tr}(B^2)$ is used. Here $F$ is the field-strength
corresponding to the gauge-field, and $B$ is an auxiliary field
in the adjoint representation of the gauge group. Carrying a path
integral over this field, leaves an effective Lagrangian for the
gauge field of the form ${\rm tr}(F^2)$ \cite{BlTh}. In a gYM$_2$,
on the other hand, one uses an arbitrary class function of the
auxiliary field $B$, instead of ${\rm tr}(B^2)$ \cite{Wit}. In
\cite{GaSoYa} the partition function and the expectation values of
the Wilson loops for gYM$_2$'s were calculated. It is worthy of
mention that for gYM$_2$'s, one can not eliminate the auxiliary
field and obtain a Lagrangian for the gauge field. One can,
however, use standard path-integration and calculate the
observables of the theory. This was done in \cite{KhAl}.

To study the behaviour of these theories for large groups is also
of interest. This was studied in \cite{Rus} and \cite{DoKa} for
ordinary YM$_2$ theories and then in \cite{AgAlKh} for YM$_2$ and
in \cite{AlKhAg} and \cite{AlTo} for gYM$_2$ theories. It was
shown that YM$_2$'s and some classes of gYM$_2$'s have a
third-order phase transition in a certain critical area.

In \cite{SK} another generalization of YM$_2$'s was introduced, and that was
to use a non-local action for the auxiliary field. There, the classical
behavior, the quantum behavior, and the large-group baehavior of the system
on a sphere, were studied.

The large-group behavior of the model on a cylinder or a disk was
investigated in \cite{GM} for YM$_2$ and in \cite{KM} for gYM$_2$. Here we
want to study the large-group behavior of a nonlocal YM$_2$ (or gYM$_2$) on
a cylinder.

The scheme of the present paper is the following. In section 2, it is shown
that the dominant representation for large-group models on a cylinder is
obtained from a generalized Hopf equation, the same Hopf equation used for
the corresponding local theory. The only difference is that the area of the
cylinder is replaced by an effective area involving the dominant
representation itself.

In section 3, the critical behavior of the model is investigated, and for
some special boundary conditions, an equation governing the critical area
corresponding to a nonlocal Yang Mills theory is obtained.

\section{The dominant representation for a large-$N$ non-local generalized
Yang-Mills theory on a cylinder}
In \cite{SK}, a non-local Yang-Mills theory was defined through
\be\label{1} e^S:=\int\rD B\;\exp\left\{\int\rd\mu\;i\;{\rm tr}(BF)+w
\left[\int\rd\mu\;\Lambda (B)\right]\right\} ,
\ee
where $F$ is the field-strength, $B$ is an auxiliary field in the
adjoint representation of the gauge group, and $\Lambda$ is a
similarity-invariant function. 
It was further shown that the wave function for this theory on a cylinder is
\be\label{2}
Z(U_1,U_2)=\sum_R \chi_R(U_1^{-1})\chi_R(U_2^{-1})\exp\{w[C_\Lambda (R)A]\} ,
\ee
where the summation runs over irreducible representations of the gauge group,
$U_1$ and $U_2$ are the path-ordered exponentials of the gauge field on the
boundaries, $\chi$ is the character of the group element, and $A$ is the
area of the surface. $C_\Lambda$ is some function related to $\Lambda$.
Taking $C_\Lambda$ a linear function of the rescaled Casimirs of the gauge
group U($N$),
\be\label{3}
\tilde C_l(R):={1\over{N^{l+1}}}\sum_{i=1}^N(n_i+N-i)^l,
\ee
where $n_i$'s are nonincreasing functions of $i$ characterizing the
representation (the Yang-tableau), one defines a function $W$ as
\be\label{4}
-N^2W\left[ A\sum_l a_l\tilde C_l(R)\right] :=w[AC_\Lambda (R)].
\ee
In the large-$N$ limit, the exponential in (\ref{2}) becomes
\be\label{5}
\exp\{w[C_\Lambda (R)A]\}=\exp\left\{ -N^2W\left[ A\int_0^1\rd x\; G(\phi )
\right]\right\},
\ee
where
\be\label{6}
G(\phi):=\sum_l(-1)^la_l\phi^l.
\ee
Also, following \cite{Rus},
\be\label{7}
\phi:={{i-n_i-N}\over N},
\ee
and
\be\label{8}
x:={i\over N}.
\ee
Following \cite{GM}, one can write the characters in (\ref{2}) as a function
of $\phi (x)$, and the eigenvalue densities
$\sigma_1(\theta)$ and $\sigma_2(\theta)$ of the boundary matrices $U_1$
and $U_2$. Then, for Large $N$, (\ref{2}) is written as 
\be\label{9}
Z=\int\rD\phi\exp\left\{ -N^2 W\left[ A\int_0^1\rd x\; G(\phi )\right] +N^2
\Gamma [\phi ,\sigma_1,\sigma_2]\right\} .
\ee
Note that the exponent in (\ref{9}) consisits of two parts. The first part
depends on both $W$ and $G$. The second part, coming from the characters,
depends on neither $W$ nor $G$. For $N\to\infty$, the wave function
(\ref{9}) is determind by the representation maximizing the exponent. This
representation satisfies
\be\label{10}
-A\; W'\left[ A\int_0^1\rd x\; G(\phi )\right] G'[\phi(x)] +
{{\delta\Gamma}\over{\delta\phi (x)}}=0.
\ee
Defining
\be\label{11}
\tilde A:= A\; W'\left[ A\int_0^1\rd x\; G(\phi )\right] ,
\ee
it is obvious that this equation is equivalent to the equation determining
the dominant representation in
\be\label{12}
\tilde Z=\int\rD\phi\exp\left\{ -N^2 \tilde A\int_0^1\rd x\; G(\phi ) +N^2
\Gamma [\phi ,\sigma_1,\sigma_2]\right\} .
\ee
But the dominant representation of this has been found in \cite{KM}.
Defining the Yang-tableau density \cite{Rus}
\be\label{13}
\rho (\phi):={{\rd x}\over{\rd\phi}},
\ee
it has been shown in \cite{KM} that in order to obtain the Yang-tableau
density corresponding to the dominant representation, one should solve the
generalized Hopf equation
\be\label{14}
{\partial\over{\partial t}}(v\pm i\pi\sigma)+{\partial\over{\partial\theta}}
G[-i(v\pm i\pi\sigma)]=0,
\ee
with the boundary conditions
\bea\label{15}
\sigma(t=0,\theta)&=&\sigma_1(\theta)\cr
\sigma(t=\tilde A,\theta)&=&\sigma_2(\theta).
\eea
Then, if there exists some $t_0$ for which
\be\label{16}
v(t_0,\sigma)=0,
\ee
one denotes the value of $\sigma$ for $t=t_0$ by $\sigma_0$:
\be\label{17}
\sigma_0(\theta):=\sigma(t_0,\theta),
\ee
and the desired density satisfies
\be\label{18}
\pi\rho[-\pi\sigma_0(\theta)]=\theta
\ee
What is shown is that from this point of view, the non-local theory behaves
like a local theory but with a surface area $\tilde A$ instead of $A$. Note,
however, that $\tilde A$ itself depends on the Yang-tableau density of the
dominant representation, through (\ref{11}) or equivalently
\be\label{19}
\tilde A=A\; W'\left[ A\int\rd z\;\rho(z)G(z)\right] .
\ee
A special case of this result was obtained in \cite{SK}, where non-local
generalized Yang-Mills theories on the sphere were studied. It was shown
there that in the limit $N\to\infty$, the theory is like a local generalized
Yang-Mills theory with the surface area $\tilde A$ instead of $A$. The
dependence of $\tilde A$ on $A$ and $\rho$ was the same as (\ref{19}).

\section{The critical behavior of the non-local Yang-Mills theory}
A non-local Yang-Mills theory is defined by
\be\label{20}
G(\phi)={1\over 2}\phi^2.
\ee
In \cite{GM}, the critical area for a Yang-Mills theory on a disk,
$\sigma_1(\theta)=\delta(\theta)$, has been found as:
\be\label{21}
A_{\rm cr}^{-1}={1\over\pi}\int{{\rd\theta'\;\sigma_2(\theta')}\over
{\pi -\theta'}}.
\ee
For a sphere, $\sigma_2(\theta)=\delta(\theta)$, and one arrives at the
familiar result
\be\label{22}
A_{\rm cr}=\pi^2.
\ee
These results can be used to obtain the critical area for a non-local
Yang-Mills theory on a disk. One can obtain $\tilde A_{\rm cr}$ as
\be\label{23}
\tilde A_{\rm cr}^{-1}={1\over\pi}\int{{\rd\theta'\;\sigma_2(\theta')}\over
{\pi -\theta'}}.
\ee
There remains, however, one problem. To obtain $A_{\rm cr}$ from
$\tilde A_{\rm cr}$, using (\ref{19}), one needs the critical density
$\rho_{\rm cr}$. Even for the disk, it is not easy to find a closed form
for $\rho_{\rm cr}$ for arbitrary $\sigma_2$. On the sphere, the situation
is better. In \cite{GM} it has been shown that the solution to the Hopf
equation for $\sigma_1(\theta)=\sigma_2(\theta)=\delta(\theta)$ is
\be\label{24}
\pi\sigma (t,\theta)={{\tilde A}\over{2t(\tilde A-t)}}\sqrt{{{4t(\tilde A-t)}
\over{\tilde A}}-\theta^2},
\ee
and
\be\label{25}
v(t,\theta)={{(2t-\tilde A)\theta}\over{2t(t-\tilde A)}}.
\ee
From this, one finds
\be\label{26}
t_0={{\tilde A}\over 2}.
\ee
Inserting this in (\ref{24}), one arrives at
\be\label{27}
\pi\sigma_0(\theta)={2\over{\tilde A}}\sqrt{\tilde A-\theta^2}.
\ee
So, using (\ref{18}),
\be\label{28}
\rho(z)={{\tilde A}\over{2\pi}}\sqrt{{4\over{\tilde A}}-z^2}.
\ee
At the critical area, the maximum of $\rho$ becomes $1$. This shows that
\be\label{29}
\tilde A_{\rm cr}=\pi^2,
\ee
as expected. But now, one can insert the critical density in (\ref{18}) to
obtain
\be\label{30}
\pi^2=A_{\rm cr}\; W'\left({{A_{\rm cr}}\over{\pi^2}}\right).
\ee
This is in accordance with what found in \cite{SK}.

One can go further. Consider a disk with the boundary condition
\be\label{300}
\sigma_2(\theta)={2\over{\pi s^2}}\sqrt{ s^2-\theta^2}.
\ee
The solution to the Hopf equation with this boundary condition is easily
obtained using the solution to the Hopf equation for the boundary conditions
corresponding to the sphere. One finds
\be\label{31}
\pi\sigma (t,\theta)={{A_0}\over{2t(A_0-t)}}\sqrt{{{4t(A_0-t)}
\over{A_0}}-\theta^2},
\ee
and
\be\label{32}
v(t,\theta)={{(2t-A_0)\theta}\over{2t(t-A_0)}},
\ee
where $A_0$ is defind through
\be\label{33}
{{4\tilde A(A_0-\tilde A)}\over{A_0}}:=s^2,
\ee
or
\be\label{34}
A_0:={{4\tilde A^2}\over{4\tilde A -s^2}}.
\ee
Again, one sets $v=0$ to obtain $\sigma_0$:
\be\label{35}
\pi\sigma_0(\theta)={2\over{A_0}}\sqrt{A_0^2-\theta^2}.
\ee
From this,
\be\label{36}
\rho (z)={{A_0}\over{2\pi}}\sqrt{{4\over{A_0}}-z^2}.
\ee
Not that for the specific boundary condition (\ref{30}), the shape of the
Yang-tableau density $\rho$ is always the semi-ellipse function obtained
for the sphere. At the critical point,
\be\label{37}
\rho_{\rm cr}(z)={\pi\over 2}\sqrt{{4\over{\pi^2}}-z^2}.
\ee
Again, this is universal, as long as the boundary condition is like
(\ref{30}). Putting this in (\ref{18}), one obtaines
\be\label{38}
\tilde A_{\rm cr}=A_{\rm cr}\; W'\left({{A_{\rm cr}}\over{\pi^2}}\right).
\ee
To find $A_{\rm cr}$, one needs $\tilde A_{\rm cr}$, which is obtained from 
(\ref{33}), and using $A_{0,\rm cr}=\pi^2$:
\be\label{39}
\tilde A_{\rm cr}=\left({{\pi^2}\over 2}\right)(1+\sqrt{1-s^2/\pi^2}).
\ee
Combining this with (\ref{38}), one arrives at
\be\label{40}
{{1+\sqrt{1-s^2/\pi^2}}\over 2}={{A_{\rm cr}}\over{\pi^2}}
W'\left({{A_{\rm cr}}\over{\pi^2}}\right).
\ee

What is achieved till now, is to obtain the critical density for the sphere
and for a disk with certain boundary conditions making the disk {\it a part
of a sphere}. The critical area can also be found for a cylinder which is
{\it a part of a sphere}. Consider the boundary conditions
\bea\label{41}
\sigma_1(\theta)&=&{2\over{\pi s_1^2}}\sqrt{ s_1^2-\theta^2}\cr
\sigma_2(\theta)&=&{2\over{\pi s_2^2}}\sqrt{ s_2^2-\theta^2}.
\eea
One can use (\ref{31}) and (\ref{32}) as the solutions to the Hopf equation,
but with (\ref{33}) replaced by
\bea\label{42}
{{4t_1(A_0-t_1)}\over{A_0}}&:=&s_1^2,\cr
{{4t_2(A_0-t_2)}\over{A_0}}&:=&s_2^2,
\eea
and
\be\label{43}
\tilde A=t_2-t_1.
\ee
Following the same arguements used for the disk, one obtains
\bea\label{44}
t_1&=&\left({{\pi^2}\over 2}\right)\left(1-\sqrt{1-s_1^2/\pi^2}\right),\cr
t_2&=&\left({{\pi^2}\over 2}\right)\left(1+\sqrt{1-s_2^2/\pi^2}\right),
\eea
and
\be\label{45}
\tilde A_{\rm cr}=\left({{\pi^2}\over 2}\right)
\left(\sqrt{1-s_2^2/\pi^2}+\sqrt{1-s_1^2/\pi^2}\right).
\ee
Using this, the critical area is found to satisfy
\be\label{46}
{{\sqrt{1-s_2^2/\pi^2}+\sqrt{1-s_1^2/\pi^2}}\over 2}=
{{A_{\rm cr}}\over{\pi^2}}W'\left({{A_{\rm cr}}\over{\pi^2}}\right).
\ee

The last thing to be considered is the the case of a disk which is {\it
almost a sphere}, that is a disk with the boundary condition
$\sigma_2(\theta)\approx\delta(\theta)$. By this approximation, it is meant
that $\sigma_2$ is an even function and one takes into account only the
second moment of $\theta$:
\be\label{47}
r:=\int\rd\theta\;\sigma_2(\theta)\theta^2.
\ee
It is assumed that $\sigma_2$ is narrowly localized around $\theta =0$, so
that one ac neglect the effect of the higher moments of $\theta$. As only
the second moment of $\theta$ is important, one can approximate $\sigma_2$
with (\ref{300}), for a small value of $s$. This value of $s$ is related to
$r$ through
\bea\label{48}
r&=&\int_{-s}^s\rd\theta\;\theta^2\left({2\over{\pi\; s^2}}
\sqrt{s^2-\theta^2}\right)\cr
&=&{{s^2}\over 4}.
\eea
One can substitute this value of $s$ in (\ref{40}) to obtain
\be\label{49}
1-{r\over{\pi^2}}={{A_{\rm cr}}\over{\pi^2}}
W'\left({{A_{\rm cr}}\over{\pi^2}}\right).
\ee
An exactly similar argument can be used for a cylinder with the boundary
conditions near a delta function. The result would be
\be\label{50}
1-{{r_1+r_2}\over{\pi^2}}={{A_{\rm cr}}\over{\pi^2}}
W'\left({{A_{\rm cr}}\over{\pi^2}}\right),
\ee
where
\be\label{51}
r_i:=\int\rd\theta\;\sigma_i(\theta)\theta^2.
\ee

Similar arguements may work for special boundary conditions and nonlocal
generalized Yang-Mills theories, provided the dominant representation
of the system is known for a sphere.

\end{document}